\documentclass[apj]{emulateapj}

\shorttitle{X-ray emission from SN 1998bw}
\shortauthors{Waxman}

\begin{document}

\title{Does the detection of X-ray emission from SN1998bw support its association with GRB980425?}

\author{Eli Waxman\altaffilmark{1}}
\altaffiltext{1}{Physics Faculty, Weizmann Institute of Science, Rehovot 76100, Israel}

\begin{abstract}

We show that the recent identification of X-ray emission from SN1998bw is naturally explained as synchrotron emission from a shock driven into the wind surrounding the progenitor by a mildly relativistic shell ejected by the supernova, the existence of which was inferred earlier from radio observations. X-ray observations imply a shell energy $E\approx10^{49.7}$~erg, and constrain the initial shell velocity $\beta c$ and normalized wind mass loss rate, $\dot{m}\equiv(\dot{M}/10^{-5}M_\odot{\rm yr}^{-1})/(v_{\rm w}/10^3{\rm km\,s^{-1}})$, to satisfy $\beta^3\dot{m}\approx10^{-1.5}$. The inferred energy is consistent with energy estimates based on radio observations provided $\dot{m}\approx0.04$, in which case radio observations imply $\beta\approx0.8$, consistent with the X-ray constraint $\beta^3\dot{m}\approx10^{-1.5}$. While X-ray observations allow to determine the parameters characterizing  the pre-explosion wind and the mildly relativistic shell ejected by SN1998bw, they do not provide evidence for existence of an off-axis "standard" GRB jet associated with SN1998bw, that may have produced GRB980425. The lack of observational signatures, typically expected to be produced by such an off-axis jet on a 1~yr time scale, may be due to the low $\dot{m}\lesssim0.1$, which implies that an off-axis jet will become observable only on $\gtrsim10$~yr time scale.

\end{abstract}


\section{Introduction}

Recent Chandra observations of SN1998bw \citep{Chandra} have allowed to identify one of the X-ray sources detected earlier by BeppoSAX \citep{Pian00} as associated with the supernova. We discuss here the implications of the detected X-ray emission, in particular to the suggested association of SN1998bw with GRB980425.

The association of GRBs with type Ib/c supernovae is motivated by the coincidence of GRB980425 and SN1998bw \citep{Galama98b}, and by the identification of a SN1998bw-like spectrum in the optical afterglow of GRB030329\footnote{See, however, \citet{Katz94}, who suggests that a SN like emission may result from the impact of the relativistic GRB debris on a nearby dense gas cloud.} \citep{Stanek03,Hjorth03}. The $\gamma$-ray luminosity $L_{\gamma,\rm Iso.}\simeq10^{51}{\rm erg/s}$ of GRB030329, inferred from the redshift z=0.1685 of its host galaxy, lies within the range of typical cosmological GRBs, $L_{\gamma,\rm Iso.}\simeq10^{52\pm1}{\rm erg/s}$ \citep[e.g. ][]{Schmidt01}. The subscript "Iso." indicates luminosity derived assuming isotropic emission. The association of GRB980425 with SN1998bw sets the distance to this burst to 38~Mpc (for $H_0=65{\rm km/s\,Mpc}$), implying that its luminosity is nearly 5 orders of magnitude lower than that typical for cosmological GRBs \citep{Pian00}.
 
Two hypotheses are commonly discussed, that may account for the orders of magnitude difference in luminosity. First, it may be that SNe Ib/c produce two different classes of GRBs, with two different characteristic luminosities. It is now commonly believed that long duration, $T>2$~s, cosmological, $L_{\gamma,\rm Iso.}\simeq10^{52}{\rm erg/s}$, GRBs are produced by the collapse of SN Ib/c progenitor stars. It is assumed that the stellar core collapses to a black-hole, which accretes mass over a long period, $\sim T$, driving a relativistic jet that penetrates the mantle/envelope and then produces the observed GRB \citep{Woosley93,Pac98,MacFadyen99}. This scenario is supported by the association of GRB030329 with SN2003dh, and by additional evidence for optical supernovae emission in several GRB afterglows \citep{Bloom03}. The origin of a second, low-luminosity, class is unknown. It may be, e.g., due to supernova shock break-out \citep{Colgate68,WoosleyWeaver86,MatznerMcKee99,Tan01}. 
 
A second possibility is that GRB980425 was a typical, cosmological GRB jet viewed off-axis \citep{Nakamura98,Eichler99,Woosley99,Granot02,Yamazaki03}. Due to the relativistic expansion of jet plasma, with Lorentz factor $\Gamma\gtrsim100$ during $\gamma$-ray emission \citep{Krolik,Baring}, $\gamma$-rays are concentrated into a cone of opening angle comparable to the jet opening angle $\theta_j$ (assuming $\theta_j>1/\Gamma$). Thus, if the jet is viewed from a direction making an angle larger than $\theta_j$ with the jet axis, the $\gamma$-ray flux may be strongly suppressed. In this scenario, strong radio emission, $L_\nu\sim10^{30}\nu_{10\rm GHz}^{-1/2}{\rm erg/s\,Hz}$, is expected at $\sim1$~yr delay \citep{FWK00,LnW00} as the jet decelerates to sub-relativistic speed and its emission becomes nearly isotropic \citep[for a more detailed discussion see][and references therein]{W04}. 

We have recently shown (Waxman 2004) that the low radio luminosity of SN1998bw, compared to that expected from a decelerated GRB jet at $\sim1$~yr delay, may be consistent with the off-axis jet interpretation of GRB980425 provided that either the magnetic field energy fraction behind the shock driven by the jet into the wind is atypically low, $\epsilon_B\le10^{-4}$, or that the density of the wind surrounding the progenitor is lower than typically expected, $\dot{m}\equiv(\dot{M}/10^{-5}M_\odot{\rm yr}^{-1})/(v_{\rm w}/10^3{\rm km\,s^{-1}})\simeq0.1$. Lower values of $\dot{m}$ and of $\epsilon_B$ reduce the specific luminosity at the transition to sub-relativistic expansion. A low value of $\dot{m}\ll1$ further delays the time at which the flow approaches sub-relativistic expansion. We consider the low $\epsilon_B$ scenario less likely, since we expect $\epsilon_B$, which is determined by the shock micro-physics, to be similar for different bursts, and $\epsilon_B$ close to equipartition is inferred from radio observations of other bursts. The latter scenario, $\dot{m}\simeq0.1$, is consistent with the constraints imposed on $\dot{m}$ by the observed radio emission from the supernova \citep{Kulkarni98}, which is interpreted as due to synchrotron emission from a shock driven into the wind by a mildly relativistic shell ejected by the supernova explosion \citep{Kulkarni98,WL99,CL99}. In this scenario, transition to sub-relativistic expansion is expected over $\sim10$~yr time scale \citep{W04}. 

The recent identification of the X-ray emission of SN1998bw was argued to support the interpretation that  GRB980425 was an off-axis "standard" GRB jet associated with SN1998bw \citep{Chandra}. We show here that the observed X-ray emission may be naturally explained as synchrotron emission from a shock driven into the wind surrounding SN1998bw progenitor by the mildly relativistic shell that was inferred to exist from earlier radio observations. In \S\ref{sec:model} we construct a simple model that describes the emission from a shock driven into the wind by a sub-relativistic shell, based on the analysis of \citet{W04}. In \S\ref{sec:98bw} we show that the model can account for both X-ray and radio observations. While X-ray observations, or radio observations, alone do not allow to determine all model parameters, combined X-ray and radio data over constrain the model, and therefore enhance our confidence in its validity. Our conclusions are summarized in \S\ref{sec:discussion}.

\section{Model}
\label{sec:model}


\subsection{Dynamics}
\label{sec:dynamics}

Let us consider a sub-relativistic shell ejected by the supernova explosion, with mass $M$, total energy $E$ and initial velocity $v=\beta c$. We assume the shell's thermal energy is dominated by its kinetic energy, since the thermal energy deposited in the shell by the passage of the supernova shock leads to expansion and post-shock acceleration which converts the thermal energy to kinetic energy \citep[As pointed out by][this post-shock acceleration is essential for achieving mildly relativistic shell velocity]{MatznerMcKee99}. The shell is assumed to propagate into a $r^{-2}$ density profile created by stellar mass loss. We assume a density profile
\begin{equation}\label{eq:rho}
    \rho=K r^{-2},\quad K\equiv\frac{\dot{M}}{4\pi v_{\rm w}},
\end{equation}
where $\dot{M}$ is the mass loss rate and $v_{\rm w}$ is the wind velocity. 

As the shell expands, it drives a shock wave into the surrounding wind, and a reverse shock is driven backward (in the shell frame) into the shell. It is straightforward to show that as long as the mass of shocked wind plasma, $M_w(r)=4\pi Kr$, is small compared to the shell's mass, the shock driven backward into the shell does not lead to significant deceleration. Denoting by $u<v$ the velocity of shocked wind (and shell) plasma, pressure balance between the forward and reverse shocks implies $\rho_w u^2=\rho_s (v-u)^2$, where $\rho_w=M_w/4\pi r^3$ and $\rho_s=\eta_s M/4\pi r^3$ denote wind and shell density respectively. Here $r/\eta_s$ is the thickness of the ejected shell. For $M_w\ll M$ we find $(v-u)/v=(M_w/\eta_s M)^{1/2}$. Thus, at small radii, where $M_w\ll M$, the shell expands with a time independent velocity, $v=\beta c$. At this stage, the velocity of the shock propagating into the wind is also time independent and given by $(\hat{\gamma}+1)v/2$, where $\hat{\gamma}$ is the adiabatic index of the wind plasma. 

Significant deceleration of the shell begins at a radius where the shocked wind mass, $M_w(r)$, is similar to the shell's mass $M$. At larger radii the flow approaches the Sedov-von Neumann-Taylor solutions describing expansion of a spherical strong shock wave into a $r^{-2}$ density profile (e.g. Chapter XII of Zel'dovich \& Raizer 2002). In these solutions, the shock radius is given by
\begin{equation}\label{eq:R}
    R=\xi(\hat{\gamma})\left(\frac{E}{K}t^2\right)^{1/3}.
\end{equation}
$\xi(\hat{\gamma})$ is a dimensionless parameter of order unity, which may be approximated by $\xi(\hat{\gamma})=(3/2)[(\hat{\gamma}+1)^2(\hat{\gamma}-1)/2\pi(7\hat{\gamma}-5)]^{1/3}$ \citep{W04}.
For $\hat{\gamma}=5/3$, appropriate for sub-relativistic flow, we have $\xi=0.73$. At the radius where $M_w(r)\sim M$, the velocity given by the Sedov-von Neumann-Taylor for the appropriate energy $E$ is similar to the initial shell velocity $v$. We therefore define the deceleration radius, $R_{\rm dec}$, as the radius at which the Sedov-von Neumann-Taylor post-shock fluid velocity equals $v$, $2\dot{R}/(\hat{\gamma}+1)=v$,
\begin{equation}\label{eq:R_dec}
    R_{\rm dec}=\frac{16\pi}{9}\xi^3\frac{E/[(\hat{\gamma}+1)v/2]^2}{\dot{M}/v_w}=
    2.1\times10^{15}\frac{E_{49}}{\beta^2\dot{m}}\,{\rm cm},
\end{equation}
where $E=10^{49}E_{49}$~erg. The deceleration time, $t_{\rm dec}$, is defined as 
\begin{equation}\label{eq:t_dec}
    t_{\rm dec}\equiv
    \frac{R_{\rm dec}}{(\hat{\gamma}+1)v/2}=0.62\frac{E_{49}}{\beta^3\dot{m}}\,{\rm day}.
\end{equation}

For $t\ll t_{\rm dec}$ we therefore expect shell expansion at constant velocity, $v$, and a shock driven into the wind at a constant speed, $(\hat{\gamma}+1)v/2$. For $t\gg t_{\rm dec}$, we expect the expansion to follow the Sedov-von Neumann-Taylor solution, Eq.~(\ref{eq:R}), with shock velocity $\dot{R}\propto t^{-1/3}$.

\subsection{Synchrotron emission}
\label{sec:synch}

Let us consider synchrotron emission from electrons accelerated to relativistic energies by the collisionless shock driven into the wind. We assume that a constant fraction $\epsilon_e$ ($\epsilon_B$) of the post-shock thermal energy is carried by relativistic electrons (magnetic field), and that the electron distribution function follows a power-law, ${\rm d}\ln n_e/{\rm d}\ln \gamma_e=-p$ for $\gamma_e\ge\gamma_m$. In what follows, we assume $p=2$, as observed for non-relativistic~\citep{non-rel} as well as for relativistic shocks~\citep{relativistic}. This power  law is produced by Fermi acceleration in collisionless shocks~\citep{non-rel,RelShock1,RelShock2}, although a first principles understanding of the process is not yet available. 

At times $t\gg t_{\rm dec}$ the flow is well described by the spherical non-relativistic self-similar solution, where the shock radius is given by Eq.~(\ref{eq:R}). For simplicity, we assume that at this stage the shocked plasma is concentrated into a thin shell behind the shock, $R/\eta\ll R,$ within which the plasma conditions are uniform (the Chernyi-Kompaneets approximation). This implies, in particular, $\eta=(\hat{\gamma}+1)/(\hat{\gamma}-1)$. The synchrotron emission obtained under the above assumptions was derived in ~\citep{W04}. The scaling of magnetic field amplitude $B$ and of $\gamma_m$ with time, at $t\gg t_{\rm dec}$, is~\citep{W04}
\begin{equation}\label{eq:Bscaling}
    B\propto t^{-1},\quad \gamma_m\propto t^{-2/3}, 
\end{equation}
and for $p=2$ the specific luminosity $L_\nu$ is
\begin{eqnarray}\label{eq:L_nu}
\nonumber  \nu L_\nu &\approx& 10^{42}(3\epsilon_e)(3\epsilon_B)^{3/4}\dot{m}^{3/4}E_{49} \\
   &\times& \left(\frac{h\nu}{1{\rm keV}}\right)^{1/2}
  \left(\frac{t}{100{\rm day}}\right)^{-3/2}\,{\rm erg/s}. 
\end{eqnarray}  
Eq.~(\ref{eq:L_nu}) holds at frequencies well above the characteristic synchrotron frequency of the lowest energy electrons ($\gamma_e=\gamma_m$), and below the cooling frequency, $\nu_c$. $\nu_c$, the characteristic synchrotron frequency of electrons for which the synchrotron cooling time is comparable to the shock expansion time, $t$, is 
\begin{equation}\label{eq:nu_c}
  \nu_c\approx2\times10^{13}(3\epsilon_B)^{-3/2}\dot{m}^{-3/2}
  \left(\frac{t}{100{\rm day}}\right)\,{\rm Hz}.
\end{equation}

Let us now consider synchrotron emission from the shock driven into the wind at times $t\ll t_{\rm dec}$. At frequencies $\nu>\nu_c$, the specific luminosity (for $p=2$) is given by $L_\nu=L_m(\nu_c/\nu_m)^{-1/2}(\nu/\nu_c)^{-1}=L_m(\nu_c\nu_m)^{1/2}/\nu$. Here, $L_m$ is the specific intensity at $\nu=\nu_m$, the characteristic synchrotron frequency of the lowest energy electrons ($\gamma_e=\gamma_m$). In order to determine the time dependence of $L_\nu$, we first note that $L_m$ is proportional to the product of the number of radiating electrons, $N_e$, and the magnetic field strength, $B$, $L_m\propto N_eB$, and that $\nu_m\propto \gamma_m^2 B$. The Lorentz factor $\gamma_c$ of electrons that cool on a time $\sim t$ is given by comparing the synchrotron cooling time, $t_{\rm syn}\propto 1/\gamma_c B^2$ to $t$, yielding $\gamma_c\propto 1/t B^2$ and $\nu_c\propto \gamma_c^2 B\propto t^{-2}B^{-3}$. The dependence of $\nu_m$, $\nu_c$, and $L_m$ on time is now obtained by noting that the time independent shock velocity implies time independent thermal energy per shocked particle, i.e. $\gamma_m\propto t^0$, thermal energy density $\propto r^{-2}\propto t^{-2}$ (since the density drops as $1/r^2$), i.e. $B\propto t^{-1}$, and $N_e\propto t$. Thus, $L_m\propto t^0$, $\nu_m\propto \gamma_m^2 B\propto t^{-1}$, $\nu_c\propto t$ and $L_\nu\propto t^0$ for $\nu>\nu_c$. Synchrotron emission at times $t\ll t_{\rm dec}$ is therefore described by
\begin{equation}\label{eq:L_early}
   \nu L_\nu\propto \left\{\begin{array}{ll}
    \nu^{1/2}t^{-1/2}, & \hbox{$\nu<\nu_c(t)$;} \\
    \nu^0 t^0, & \hbox{$\nu>\nu_c(t)$,} \\
\end{array}%
\right.    
\end{equation}
with $\nu_c$ given by Eq.~(\ref{eq:nu_c}). Combined with Eqs.~(\ref{eq:t_dec}) and~(\ref{eq:L_nu}), this determines the synchrotron emission predicted by the model at $t<t_{\rm dec}$.

The following line of arguments shows that synchrotron emission from the shock driven backward into the expanding shell, which exists at $t<t_{\rm dec}$, is dominated by the emission from the shock driven into the wind, under the assumption that $\epsilon_e$ and $\epsilon_B$ are similar for the post-shock plasma behind the two shocks. Since the thermal energy density is the same in the shocked wind and shell plasmas (pressure equilibrium), the magnetic field is the same in the two regions (assuming similar $\epsilon_B$ for both shocks) and the product of the electron number density and their minimum Lorentz factor, $n_e\gamma_m$, which is proportional to the thermal energy density, is the same in the shocked wind and shell plasmas. Similar $B$ implies similar $\nu_c$ for both shocks, and hence the ratio of the specific synchrotron luminosity of the shocked shell plasma and shocked wind plasma, $L_{\nu,s}/L_\nu$, is $L_{\nu,s}/L_\nu=L_{m,s}\nu_{m,s}^{1/2}/L_m\nu_m^{1/2}=N_{e,s}\gamma_{m,s}/N_e\gamma_m=\Delta_s/\Delta$, where the subscript $s$ denotes quantities related to the shocked shell plasma, and $\Delta$ ($\Delta_s$) is the width of shock wind (shell) plasma. The thickness of the region of shocked wind/shell plasma is proportional to the velocity of the shock propagating into the wind/shell plasma, measured in the shocked plasma rest frame, $\Delta_s/\Delta=(v-u)/u$ [see discussion following Eq.~(\ref{eq:rho})]. Thus, for $M_w\ll M$ we find $L_{\nu,s}/L_\nu=(M_w/\eta_sM)^{1/2}\ll1$.

The following point should be emphasized here. The model described above is highly simplified. For example, the wind density is assumed to follow a pure $1/r^2$ law and to be free of inhomogeneities, and the expanding shell is assumed uniform with no internal (density, velocity) gradients. In comparing model predictions with observations one should therefore expect an approximate, rather than exact, agreement. Construction of a more detailed, and more accurate, model requires a more detailed knowledge of, for example, the wind density distribution and its homogeneities and of the internal structure of the shell. We find that such a more detailed analysis is not warranted by the present data, since the number of free parameters in a more detailed model would be larger than the observational constraints.

\section{Implications to SN1998bw}
\label{sec:98bw}

The luminosity of the X-ray source associated with SN1998bw, given in Fig.~3 of \citet{Chandra}, may be approximately described as
\begin{equation}\label{eq:L_obs}
    L_X\approx 6\times10^{40}{\rm erg/s}\,\times\left\{%
\begin{array}{ll}
    1, & \hbox{$t<$110 day;} \\
    (t/110{\rm day})^{-3/2}, & \hbox{otherwise.} \\
\end{array}%
\right.    
\end{equation}
The $t^{-3/2}$ decline at late times is consistent with the prediction of Eq.~(\ref{eq:L_nu}), describing synchrotron emission from the shock driven into the wind at the self-similar stage of expansion, $t>t_{\rm dec}$ [see Eq.~(\ref{eq:t_dec})], at frequencies lower than the cooling frequency, given by Eq.~(\ref{eq:nu_c}). The time independent flux at earlier time is consistent with synchrotron emission from the shock driven into the wind at the stage of expansion at constant speed, $t<t_{\rm dec}$, at frequencies higher than the cooling frequency [see Eq.~(\ref{eq:L_early})]. The observed X-ray luminosity may therefore be interpreted as due to the shock driven by a sub-relativistic shell into the wind, for which deceleration starts at $t_{\rm dec}\approx110$~day and the cooling frequency passes through the X-ray band at $t\sim t_{\rm dec}$, i.e. $h\nu_c(t=110\,{\rm day})\approx1$~keV. 

Using Eq.~(\ref{eq:nu_c}), the requirement $h\nu_c(t=110\,{\rm day})\sim1$~keV implies
\begin{equation}\label{eq:nu_c_con}
  (3\epsilon_B\dot{m})^{3/4}\approx 10^{-2}.
\end{equation}
Using this result in Eq.~(\ref{eq:L_nu}), the observed luminosity at $t=t_{\rm dec}\approx110$ implies
\begin{equation}\label{eq:L_con}
  3\epsilon_e E_{49}\approx 6.
\end{equation}
Using this result in Eq.~(\ref{eq:t_dec}), the requirement $t_{\rm dec}\approx110$~day implies
\begin{equation}\label{eq:t_con}
  \beta^3\dot{m}\approx 0.03/(3\epsilon_e).
\end{equation}

Our model for the X-ray emission of SN1998bw is consistent with its radio emission, which is also interpreted as synchrotron emission from a shock wave driven into the wind by a mildly relativistic shell ejected by the supernova explosion. Within the framework of this model, radio observations imply electron energy fraction close to equipartition, $3\epsilon_e\sim1$ \citep[see Eq.~(17) of][]{WL99}. Near equipartition value of $\epsilon_e$ is also inferred from radio observations of GRB970508 for the shock driven by the fireball into surrounding plasma at late stages, where the shock becomes mildly relativistic \citep{FWK00}. We therefore adopt $3\epsilon_e\sim1$, which implies, using Eq.~(\ref{eq:L_con}), $E_{49}\approx6$. 

Radio observations further allow to determine the values of $\epsilon_B$, $\beta$ and $E$ as functions of $\dot{m}$. The lowest allowed value of $\dot{m}$ (and of $E$) is that corresponding to magnetic field equipartition, with larger values of $\dot{m}$ implying lower values of $\epsilon_B$ (and larger values of $E$). Model predictions were shown \citep{Kulkarni98,WL99,CL99} to be consistent with radio observations for, e.g., \{$\dot{m}=0.04$, $\epsilon_B\approx0.1$, $E_{49}\approx3$, $\beta\approx0.8$\}, and for \{$\dot{m}=6$, $E_{49}\approx50$, $\epsilon_B\approx10^{-6}$, $\beta\approx0.6$\}. The energy inferred from X-ray observations, Eq.~(\ref{eq:L_con}), implies that the former choice of parameters, \{$\dot{m}=0.04$, $\epsilon_B\approx0.1$, $E_{49}\approx3$\} with $\beta\approx0.8$, should be chosen for consistency with observations. 

X-ray and radio observations therefore over constrain the model: All model parameters are determined from radio observations with the additional X-ray constraint given by Eq.~(\ref{eq:L_con}). The inferred parameter values are consistent with the two additional independent X-ray constraints, Eqs.~(\ref{eq:nu_c_con}) and (\ref{eq:t_con}).

\section{Discussion}
\label{sec:discussion}

Simple analytic expressions are given in \S~\ref{sec:model} for the specific luminosity emitted by a shock wave driven into a wind surrounding a supernova progenitor by a sub-relativistic shell ejected by the supernova explosion. We have shown in \S~\ref{sec:98bw} that the observed X-ray and radio emission from SN1998bw are consistent with such a model. In particular, the approximately time independent X-ray flux at early time, $t<100$~day, and the $t^{-3/2}$ decline of the flux at late time [see Eq.~(\ref{eq:L_obs})] are characteristic for the emission from a shock driven into a wind by a shell which suffers significant deceleration at $t\gtrsim100$~day [see Eqs.~(\ref{eq:L_nu}),~(\ref{eq:L_early})]. Combined X-ray and radio data allow to determine all model parameters: The shell kinetic energy is $E\approx10^{49.7}$~erg and its initial velocity is $\beta\approx0.8$; The normalized mass loss rate of the wind, $\dot{m}\equiv(\dot{M}/10^{-5}M_\odot{\rm yr}^{-1})/(v_{\rm w}/10^3{\rm km\,s^{-1}})$, is $\dot{m}\approx0.04$; The post-shock magnetic field is not far below equipartition, $\epsilon_B\approx0.1$. Combined X-ray and radio data over constrain the model, and therefore enhance our confidence in its validity. We note, that the inferred energy in mildly relativistic ejecta, $\Gamma\beta>1$ where $\Gamma\equiv(1-\beta^2)^{-1/2}$, is an order of magnitude higher than the value obtained from analyses of shock emergence from the assumed progenitor of SN1998bw, $\simeq10^{48.5}$~erg \citep{Tan01}. 

X-ray observations do not provide therefore evidence for existence of an off-axis "standard" GRB jet associated with SN1998bw, that may have produced GRB980425. The lack of observational signatures typically expected to be produced by such an off-axis jet on a 1~yr time scale may be due, however, to the low value of $\dot{m}$, $\dot{m}\lesssim0.1$, which implies that an off-axis jet will become observable only on $\gtrsim10$~yr time scale \citep{W04}.

\acknowledgements
This research was supported in part by AEC, Minerva, and ISF grants.

\end{document}